\documentclass[english,superscriptaddress,prl,twocolumn]{revtex4-1}
\usepackage[toc,page,title,titletoc,header]{appendix}
\usepackage[T1]{fontenc}
\usepackage{enumerate}
\usepackage{xcolor}
\usepackage{graphicx}
\usepackage[latin1]{inputenc}
\usepackage{epstopdf}
\usepackage{amsthm}

\usepackage{amsmath}
\usepackage{shapepar}
\usepackage{tabularx}
\usepackage{color}
\usepackage[colorlinks=true,linkcolor=blue,citecolor=magenta,urlcolor=blue]{hyperref}
\usepackage{amssymb, amsbsy, amsthm}
\makeatletter
\newcommand{\tr}{\mbox{\rm Tr}}

\newtheorem{Theorem}{Theorem}

\usepackage{babel}
\makeatother

\begin{document}
\title{Specifying the Intrinsic Back-action of a General Measurement}
\author{Liang-Liang Sun}
\email{sun18@ustc.edu.cn}
\affiliation{Department of Modern Physics and National Laboratory for Physical Sciences at Microscale, University of Science and Technology of China, Hefei, Anhui 230026, China}

\author{Armin Tavakoli}
\affiliation{Physics Department, Lund University, Box 118, 22100 Lund, Sweden}

\author{Ren\'{e} Schwonnek}  
\affiliation{Institut f\"{u}r Theoretische Physik, Leibniz Universit\"{a}t Hannover, Appelstrasse 2, 30167, Germany}

\author{ Matthias Kleinmann}
\affiliation{Naturwissenschaftlich-Technische Fakult\"{a}t, Universit\"{a}t Siegen, Walter-Flex-Stra\ss e 3, 57068 Siegen, Germany}

\author{Zhen-Peng Xu}
\affiliation{School of Physics and Optoelectronics Engineering, Anhui University, Hefei 230601, People's Republic of China}

\author{Sixia Yu }\email{yusixia@ustc.edu.cn} 
\affiliation{Department of Modern Physics and National Laboratory for Physical Sciences at Microscale, University of Science and Technology of China, Hefei, Anhui 230026, China}

\date{\today{}}
\begin{abstract}

Understanding the invasive nature of quantum measurement and {its} implications in quantum foundations and information science demands a mathematically rigorous and physically well-grounded characterization of intrinsic back-action in general measurement processes. However, such a framework remains elusive, leaving a critical gap in quantum theory. Here, we address this issue by conceptualizing a general quantum measurement as a reduction of extended projection measurements ensured by Naimark's theorem and, derive a state-updating rule  for the concerned measurement as a reduction of the projective measurements postulate. Our framework provides a detailed analysis by explicitly decomposing the disturbance effects into two distinct contributions: those arising from the measurement elements themselves and those resulting from the dilation process. Notably, this formulation naturally recovers the projection postulate in the case of projective measurements.  Beyond providing insights into joint measurability, non-disturbance, our rule establishes quantitative connections between intrinsic disturbance and other fundamental quantum features, such as randomness, uncertainty, and information gain.




\end{abstract}
\maketitle

Previously, a measurement has been perceived as a passive recording of pre-existing values of physical quantities assuming a realistic interpretation. This classical intuition was revolutionized by quantum theory, where a particle can exist in a superposition of multiple states, and a projection measurement forces it to randomly and abruptly "jump" into one eigenstate of the measured observable, meanwhile assigning a  value to the observable. This state change is referred to intrinsic disturbance effect of quantum measurement~\cite{book, peres1995quantum, bush2016} that underpins nearly all fundamental quantum properties~\cite{RevModPhys.95.011003}, from the initial complementarity principle to nonlocality~\cite{RevModPhys.86.419, PhysRevA.73.012112}, contextuality~\cite{045007}, steering~\cite{RevModPhys.92.015001, Chen2017}, and coherence~\cite{PhysRevA.97.062308, PhysRevA.106.042428}. It also leads to a fundamental distinction between the quantum and the classical information science by limiting the amount of information an adversary can extract from a single particle~\cite{Park1970} and ensuring that any such attempt inevitably causes a detectable disturbance,  providing  the basis for secure quantum cryptography~\cite{Wootters1982, DIEKS1982271}.  However, what is surprising is that an accurate characterization of this effect for general measurement process remains elusive, leaving significant gaps in the theoretical framework. Thus, despite recognizing certain notions---such as genuine randomness~\cite{RevModPhys.89.015004, Ma2016} from wave function collapse, information gain quantified by entropy reduction~\cite{Groenewold1971} or operationally invariant measures~\cite{PhysRevLett.83.3354}, and measurement uncertainty---as relevant to intrinsic disturbance, a precise quantitative connection between them remains elusive. And also,  the disturbance effect fundamentally limits the performance of quantum information tasks, particularly those involving sequential measurement schemes~\cite{Silva2015, Brown2020, Steffinlongo2022, Mohan2019, Miklin2020};  see the review \cite{Cai2025}, and such a limitation is not well-understood.

Traditional measurements are described as projections onto a basis, with post-measurement states corresponding to the basis elements. This is formalized in the projection measurement postulate~\cite{ von2018mathematical, 20065180904} . Gleason's theorem~\cite{gle, PhysRevLett.91.120403} provides a physical interpretation for this postulate, demonstrating that the postulate emerges from a set of operational assumptions~\cite{Cassinelli1983}. General measurements, however, are characterized by positive operator-valued measures (POVMs)~\cite{peres1995quantum}, which can be realized through Naimark's dilation to  higher-dimensional projection-valued measures (PVMs). For a given POVM, the dilation is highly non-unique, as are the instruments describing the measurement process and the associated state changes represented using the Kraus form~\cite{ref1} or Choi-Jamiolkowski  isomorphism~\cite{Choi1975, Jamiokowski1972}.  Therein, the state updating rules (SURs) defined do not explicitly specify the intrinsic disturbance and contain excessive additional degrees of freedom specific to measurement realizations, making it impractical to handle. Although L\"uders' instrument~\cite{20065180904} has led to significant results, such as the gentle measurement lemma~\cite{Winter1999} and applications in specific contexts\cite{PRXQuantum, 10113, Chia20, PhysRevA.106.L010601, 10.1145/3313276.3316378, watts2022quantum, PRXQuantum.5.020367, PhysRevLett.129.160503, PhysRevX.13.041001, PhysRevLett.129.250504},   it is not suited for many real measurement processes such as Ivanovic-Dieks-Peres (IDP) measurement\cite{IVANOVIC1987257, DIEKS1988303, PERES198819}.





By interpreting POVMs as reductions of their dilated PVMs, we derive a SUR to characterize the intrinsic measurement backaction in general measurement processes. Our SUR, following from the postulate of PVM,  comprises two distinct components: one formulated entirely in terms of measurement elements, and the other dependent on the extended PVMs, which respectively account for the contributions arising from the measurement itself and the specific experimental realization. This framework provides also a consistent understanding of the concepts of joint measurability and non-disturbance. It also enables us to establish connections between intrinsic disturbance, as defined by our SUR, and other fundamental properties. Specifically, we present: (1) a generalized version of Winter's gentle measurement lemma, which bounds state changes in terms of outcome distributions, with Winter's original result emerging as a special case; and (2) a detailed analysis of how intrinsic disturbance relates to genuine randomness, measurement uncertainty, and information gain quantified through operationally invariant measures and entropy reduction. Our SUR thus offers a comprehensive framework for understanding intrinsic disturbance, serving as a powerful tool to elucidate concepts central to quantum measurement and holds significant potential for practical applications in quantum information science.

\section{I. State updating rules for Naimark's theorem}
An  instrument ~\cite{nielsen2001quantum, bush2016, 10000}  is a description of how a given measurement  ${\mathbf{M}}=\{{\rm  {M}}_{i}\}_{i}$ is implemented on a  state of interest $\rho_{s}$ and how the state is changed. It gives   the classical outcome  $i$ with probability  $p_{i}\equiv {\rm Tr}(\rho_{s}{\rm  {M}}_{i})$ and output  the  post-measurement state $\tilde{\mathcal{J}}_{i}(\rho_{s})\equiv \mathcal{J}_{i}(\rho_{s})/p_{i}$ by  defining a SUR $\mathcal{J}\equiv \{\mathcal J_i(\cdot)\}_{i}$, where  the sum $\mathcal{J}(\cdot )\equiv \sum_{i}\mathcal{J}_{i}(\cdot)$ is a trace-preserving completely positive linear map describing the average post-measurement state.      The postulate of PVM  provides  an  immediate example  of instrument for a PVM ${\bf Q}\equiv \{{\rm  {Q}}_{i}\}_{i}$  as  ~\cite{ von2018mathematical, 20065180904} 
\begin{equation}
\mathcal {J}^{\mathcal{P}}_{i}(\cdot): \rho_s \mapsto 
{\rm  {Q}}_{i}\rho_{s} {\rm  {Q}}_{i}.
\end{equation}
This SUR  is referred to as the Neumann-L\"uders projection measurement postulate.

To implement a non-projective POVM, one has to  extend it to a PVM via  Naimark's dilation~\cite{naimark1940spekral, nielsen2001quantum}. The non-uniqueness of the possible dilations  results in various SURs, which  can be expressed in a unified form   \cite{bush2016}  
\begin{equation}
\mathcal {J}_{i}(\cdot): \rho_s \mapsto 
{\rm Tr}_{a}({\rm U}_{i}{\rm  {Q}}_{i}\rho_{s} \otimes \rho_{a}{\rm  {Q}}_{i} {\rm U}_{i}^{\dagger}),  \label{general}
\end{equation}
where $\bf Q$ is a PVM that dilates $\bf M$,  $\rho_{a}$ is the state of the ancillary system,  ${\rm U}_i$ is an unitary matrix acting  on the joint state of the  extended system $\rho_{s}\otimes \rho_{a}$,   and the trace is taken over the ancilla system. %
For latter use, we introduce notions $\rho_{sa}\equiv \rho_{s}\otimes \rho_{a}$,  $\rho_{sa|u}\equiv {\rm U}\rho_{sa}{\rm U}^{\dagger}$,   and ${\rm Q}_{i|{\rm H}}\equiv {\rm U}{\rm Q}_{i}{\rm U}^{\dagger}$ where ${\rm U}$ is unitary operation coupling $\rho_{s}$ and $\rho_{a}$.  Notably,  it follows the expression  Eq.(\ref{general}) that ${\rm Tr}(\rho_{s}{\rm M}_{i})={\rm Tr}(\rho_{sa}{\rm Q}_{i})$ $\forall \rho_{s} $ and    ${\rm M}_{i}={\rm Tr}_{a}({\rm Q}_{i}\rho_{a}\otimes \openone_{s})$, meaning  that a POVM can be interpreted as a reduction of some PVM. Interestingly,  an apparatus  implementing  a fixed PVM $\mathbf{Q}$ can realize different POVMs on $\rho_{s}$  by varying the input state  $\rho_{a}$ in the expression $\{{\rm Tr}_{a}({\rm Q}_{i} \rho_{a}\otimes {\openone}_{s})\}$. 

In Eq.(\ref{general}),  the unitary operation ${\rm U}_i$ causes the {\it reversible} state change that can be  remedied by applying a compensation operation ${\rm U}^\dagger_i$. The state change   $\rho_{sa} \mapsto {\rm Q}_{i}\rho_{sa}{\rm Q}_{i}$  is {\it irreversible} that is the origin of intrinsic disturbance on the measured state.   To characterize  the intrinsic state change of the measured system we  define a SUR   as a reduction of PVM postulate  as
\begin{equation}
\mathcal {J}^{\mathcal{N}}_{i}(\cdot): \rho_s \mapsto 
{\rm Tr}_{a}({\rm  {Q}}_{i}\rho_{sa}{\rm  {Q}}_{i}).  \label{intrinsic}
\end{equation}
 It is clear that $\mathcal{J}^{\mathcal{N}}$ is determined by the measurement $\bf M$ performed and the extended measurement  $\bf Q$ employed.     To account for how these two ingredients  affects the $\mathcal{J}^{\mathcal{N}}$, we provide an explicit and universal expression for  $\mathcal{J}^{\mathcal{N}}$ (see appendix for the proof).



\begin{Theorem}[State updating rule] For an arbitrary dilation of a given POVM, i.e.,  ${\rm M}_{i}={\rm Tr}_{a}({\rm Q}_{i}\rho_{a}\otimes \openone_{s})$, the intrinsic instrument Eq.(\ref{intrinsic}) has the following operator-sum representation
\begin{equation}\label{sur}
\mathcal {J}^{\mathcal{N}}_{i}(\cdot): \rho_s \mapsto 
\textstyle {\rm  {M}}_{i}\rho_{s}{\rm{M}}_{i}+\sum_{l}{\rm N}_{l|i}\rho_{s}{\rm N}_{l|i}^{\dagger},
\end{equation}
where $\{{\rm N}_{i|l}\}_{i,l}$ is a set of operators satisfying the condition
\begin{equation}\label{cond}
\sum_{l \neq 0} {\rm N}_{i|l}^{\dagger} {\rm N}_{j|l} = \delta_{ij} {\rm M}_{i} - {\rm M}_{i} {\rm M}_{j}.
\end{equation}
Conversely, for any set of operators $\{{\rm N}_{i|l}\}_{i,l}$ that satisfies Eq.~\eqref{cond}, there exists a Naimark's dilation, i.e.,  an ancillary state $\rho_{a}$ and a PVM $\mathbf{Q}$ on the extended system such that ${\rm Tr}_{a}({\rm Q}_{i} \rho_{a}\otimes \openone_{s}) = {\rm M}_{i}$.
\end{Theorem}
This theorem provides a necessary and sufficient characterization for the instrument expressed in Eq.(\ref{intrinsic}).  We note that  the first term of   Eq.(\ref{sur}) is fixed for a given POVM.  The second term, capturing the disturbance effect attributed to the extended $\bf Q$,  carries a weight of ${\rm Tr}(\sum_{l}{\rm N}_{l|i}\rho_{s}{\rm N}_{l|i}^{\dagger})={\rm Tr}[\rho_{s}({\rm M}_{i} - {\rm M}^{2}_{i})]$, which diminishes as ${\rm M}_{i}$ approaches a projection, thereby  the dependency of $\mathcal{J}^{\mathcal{N}}$ on the specific dilation reduces and $\mathcal{J}^{\mathcal{N}}$ recovers the Neumann-L\"uders projection postulate for projective measurements.  Our SUR directly applies to the Ivanovic-Dieks-Peres (IDP) measurement (see the appendix for details)~\cite{IVANOVIC1987257, DIEKS1988303, PERES198819}, a well-known framework for quantum state discrimination that has been experimentally validated~\cite{PhysRevLett.124.080401}.  By these facts, Eq.(\ref{sur}) thus can be seen as a natural generalization of the PVM postulate to the case of POVMs.

Here, we incorporate  the instrument provided in the text-book  in the above framework, where  a measurement is modeled as coupling a measured state $\rho_{s}$ with an ancillary system in state $\rho_{a}$ with unitary operation $\rm U$, which leads to a measured state $\rho_{sa|u}$. Performing a projection measurement $\{{\rm Q}_{i}\equiv \openone_{s}\otimes |i\rangle \langle i|_{a}\}_{i}$ on the joint system  leads to a SUR as specified by $\mathcal{J}^{\mathcal{D}}: \rho_{s}\mapsto {\rm U}_{i|s}\sqrt{{\rm M}_{i}}\rho_{s} \sqrt{{\rm M}_{i}}{\rm U}^{\dagger}_{i|s}$  where $|i\rangle \langle i|_{a}$ is a projector on the ancillary system and ${\rm U}^{\dagger}_{i|s}$ is a unitary matrix acting on the measured state determined by $\rm U$. 
This SUR can be re-expressed  as ${\rm Tr}_{a}({\rm Q}_{i}\rho_{sa|u}{\rm Q}_{i})={\rm Tr}_{a}({\rm U}^{\dagger}{\rm Q}_{i|{\rm H}}\rho_{sa}{\rm Q}_{i|{\rm H}}{\rm U})$ that  assumes  the form of Eq.(\ref{general}),  where $\{{\rm Q}_{i|{\rm H}}\equiv  {\rm U}^{\dagger}{\rm Q}_{i}{\rm U}\}_{i}$ is the measurement $\mathbf{Q}$ seen  from a Heisenberg picture  (see the appendix for detailed discussion).


By the above framework,   a POVM is the reduction of a joint PVM and SUR $\mathcal{J}^{\mathcal{N}}$ is the reduction of the postulate of PVM. This motivates us to revisit notions of non-disturbance and joint measurability while taking  the ancilla into consideration in the sequential measurement schemes. This  offer different insights as the ancilla retains  information about the measured system that would otherwise be lost in the previous consideration where the ancillary system is often discarded or initialized to some state in each round of the experiment.    Let us begin by recalling  definitions~\cite{RevModPhys.86.1261}: two measurements $A \equiv \{{\rm M}_{i}\}_{i}$ and $B \equiv \{{\rm M}'_{j}\}_{j}$ are said to be joint-measurable if there exists a joint measurement $\{{\rm M}_{ij}\}_{i,j}$ such that $\sum_{i} {\rm M}_{ij} = {\rm M}'_{j}$ and $\sum_{j} {\rm M}_{ij} = {\rm M}_{i}$. Furthermore, measuring $A$ is said to cause no disturbance in $B$ if there exists an instrument of $A$, denoted by $\mathcal{J}_{A}$, such that ${\rm Tr}(\mathcal{J}_{A}(\rho_{s}) {\rm M}'_{j}) = {\rm Tr}(\rho_{s} {\rm M}'_{j})$ for all $\rho_{s}$ and $j$. These notions are equivalent for PVMs~\cite{10063}. For POVMs, it has been established that joint-measurability $\geq$ non-disturbance~\cite{10063}. However,  here we incorporate the ancillary system and define a generalization of non-disturbance as the existence of a measurement $\mathbf{Q}_{A}$ performed on $\rho_{sa}$ to realize $A$, along with $\mathbf{Q}_{B} = \{ {\rm Q}_{j|B}\}_{j}$, such that ${\rm Tr}({\rm Q}_{j|B} \mathcal{J}^{\mathcal{P}}_{A}(\rho_{sa})) = {\rm Tr}(\rho_{s} {\rm M}'_{j}), \quad \forall \rho_{s}, j,$
where $\mathcal{J}_{A}^{\mathcal{P}}$ is the SUR associated with $\mathbf{Q}_{A}$. This generalization ensures that joint measurability is equivalent to non-disturbance. Specifically, if there exists a joint measurement $\{{\rm M}_{ij}\}_{i, j}$ with an extension $\{{\rm Q}_{ij}\}_{i, j}$, one can construct non-disturbing extensions $\mathbf{Q}_{A} = \sum_{j} {\rm Q}_{ij}$ and $\mathbf{Q}_{B} = \sum_{i} {\rm Q}_{ij}$ to realize the two POVMs. This equivalence recovers the results presented in~\cite{PhysRevA.91.022110}\footnote{To understand the existence of non-commuting measurements that are jointly measurable, we note that while some jointly measurable PVMs ${\rm Q}_{i}$ and ${\rm Q}_{j}$ are commutative, their reductions ${\rm Tr}_{a}({\rm Q}_{i(j)} \rho_{a})$ (i.e., ${\rm M}_{i}$ and ${\rm M}'_{j}$) are not necessarily commutative}. We leave also a conceptual comparison between PVM and POVM in appendix.

\section{II. Relations between disturbance and other notions}


We are in a position to explore the intricate relationship between intrinsic disturbance defined by $\mathcal{J}^{\mathcal{N}}$ and other fundamental properties, including uncertainty quantified by the $\alpha$-R\'{e}nyi entropy $\mathrm{H}_{\alpha}(\mathbf{P})$ with $\mathbf{P}\equiv \{p_{i}\}_{i}$ specifying  the statistics of the measurement, genuine randomness specified by $\mathcal{R}$, and information gain characterized by operationally invariant information $\mathcal{I}(\mathbf{P})$~\cite{PhysRevLett.83.3354}, as well as the reduction of entropy $\mathcal{G}^{\mathcal{D}}$~\cite{Groenewold1971}. Detailed definitions of these quantities will be provided later. Disturbance can be quantified by a distance measure  between the original state $\rho_{s}$ and the post-measurement state $\mathcal{J}^{\mathcal{N}}(\rho_{s})$. Below, we summarize these findings as  follows (proofs  are left in the appendix)
\begin{itemize}
    \item \textbf{Randomness-Disturbance Balance }: In the implementation of a POVM $\bf M$ on state $\rho_{s}$, the randomness $\mathcal{R}$ generated in this process  can be lower-bounded with intrinsic disturbance employing the relative entropy  $S(\rho\| \sigma)  \equiv{\rm Tr}[\rho \log(\rho -\log \sigma)]$ as 
    \begin{eqnarray}
\mathcal{R}\geq S(\rho_{s}\|\mathcal{J}^{\mathcal{N}}(\rho_{s})) \label{rand}
	\end{eqnarray}
    
    \item \textbf{Generalized Gentle Measurement Lemma}:  For an outcome \( i \), the disturbance effect can be lower-bounded by the probability of that outcome, denoted as \( p_{i} \), as follows:
    \begin{eqnarray}
p^{\alpha}_{i} \leq 	\operatorname{Tr}
	{\Large(}(\tilde{\mathcal{J}}^{\mathcal{N}}_{i})^{\frac{1-\alpha}{2\alpha}}\rho_{s}
	(\tilde{\mathcal{J}}^{\mathcal{N}}_{i})^{\frac{1-\alpha}{2\alpha}}{\Large)}^{\alpha}, \alpha\in [1/2, \infty).
	\end{eqnarray}
 
    \item \textbf{Uncertainty-Disturbance Trade-off}: The uncertainty of measurement implies an upper bound of   R\'{e}nyi entropy  ${\rm H}_{\alpha}({\bf P})=\frac{\alpha}{1-\alpha}\log \|\mathbf{P}\|_{\alpha}$  on its disturbance, employing the  $\alpha-$R\'{e}nyi divergence \( \mathcal{D}_\alpha(\rho \| \sigma) \)~\cite{Renyi1961}. Thus, we have :
\begin{eqnarray}
	{\rm H}_{\frac{1}{\alpha}}(\mathbf{P}) \geq \mathcal{D}_{\alpha}(\rho_{s}\|\mathcal{J}^{\mathcal{N}}(\rho_{s})),  \forall \alpha\in [1/2, 1). \label{UDR1}
	\end{eqnarray}
    \item \textbf{Information-Disturbance Trade-off}: The uncertainty of measurement, in terms of operationally invariant information \( \mathcal{I}(\mathbf{P}) \equiv 1 - \|\mathbf{P}\|_{2}^{2} \)~\cite{PhysRevLett.83.3354}, is lower-bounded by the trace distance $\mathcal{D}_{\rm tr}(\rho, \sigma)=1/2{\rm Tr}|\rho-\sigma|$ with  ${\rm Tr}|A|={\rm Tr}\sqrt{AA^{\dagger}}$ between the initial state \( \rho_{s} \) and the post-measurement state \( \mathcal{J}^{\mathcal{N}}(\rho_{s}) \).
\begin{eqnarray}
\mathcal{I}({\bf{P}})\geq \mathcal{D}^{2}_{\rm Tr}(\rho_{s}, \mathcal{J}^{\mathcal{N}}(\rho_{s})). \label{invari}
  \end{eqnarray}

    \item \textbf{Information-Disturbance Balance}: The randomness of measurement and the resulting disturbance exhibit a competitive relationship:
\begin{eqnarray}
 \mathcal{G}^{\mathcal{D}} +\mathcal{R} = {\rm H}(\mathbf{P}),\label{compet}
  \end{eqnarray}
where  ${\rm H}({\bf {P}})\equiv -\sum_{i}p_{i}\log p_{i}$ is the Shannon entropy.  
\end{itemize}
Quantum randomness, or called genuine randomness,   manifests itself alongside the intrinsic disturbance caused by the ``random jump'' of the measured state during the measurement process. Genuine randomness is an essential resource for quantum communications, and numerous protocols have been developed to verify quantum randomness by leveraging quantum phenomena that exhibit this characteristic~\cite{RevModPhys.89.015004, Ma2016}.  For a projection measurement \(\mathbf{Q}\) performed on a state \(\rho\), the generated randomness can be quantified as $\mathcal{R} \equiv S(\rho \| \sum_{i} {\rm Q}_{i} \rho {\rm Q}_{i})$~\cite{Yuan2019}.
To implement a general POVM, one can perform an extended PVM \(\mathbf{Q}\) on the extended state described by \(\rho_{sa|\mathbf{Q}}\). The resulting randomness is   $\mathcal{R}= S(\rho_{sa|\mathbf{Q}} \| \sum_{i} {\rm Q}_{i} \rho_{sa|\mathbf{Q}} {\rm Q}_{i}),$ which is lower-bounded as
$\mathcal{R}\geq S(\rho_{s} \| \mathcal{J}^{\mathcal{N}}(\rho_{s}))$.
 To validate this lower-bound, let us consider the instrument $\mathcal{J}^{\mathcal{D}}$, where PVM $\mathbf{Q}$ is performed on $\rho_{sa|u}$, we then have   $S({\rm U}\rho_{sa}{\rm U}^{\dagger}\|\sum_{i}{\rm Q}_{i}{\rm U}\rho_{sa}{\rm U}^{\dagger} {\rm Q}_{i})=S(\rho_{sa}\|\sum_{i}{\rm Q}_{i|{\rm H}}\rho_{sa}{\rm Q}_{i|{\rm H}})\geq S(\rho_{s}\|\mathcal{J}^{\mathcal{N}}(\rho_{s}))$, where we have used the unitary invariant of relative entropy in the inequality and  data processing inequality in the inequality.    This relation immediately leads to a quantum randomness generation protocol: by independently performing an measurement on $\rho_{s}$ and $\mathcal{J}^{\mathcal{N}}(\rho_{s})$,  two probability distributions  denoted as $\mathbf{P}_{B}\equiv \{p_{i|B}\}_{i} $ and  $\mathbf{P}'_{B}\equiv \{p'_{i|B}\}_{i} $ are obtained  that can bound the randomness generation as  $\mathcal{R}\geq S(\rho_{s}\| \mathcal{J}^{\mathcal{N}}(\rho_{s})) \geq S(\mathbf{P}_{B}\|\mathbf{P}'_{B})$, with classical relative entropy  $\mathrm{H}(\mathbf{P}_{B} \| \mathbf{P}'_{B}) \equiv \sum_{i} p_{i|B} \log_{2} \frac{p_{i|B}}{p'_{i|B}}$.


Gentle measurement lemma is initially introduced specific to  L\"uders' instrument $\mathcal{J}^{{\mathcal{D}_{L}}}_{i}(\cdot): \rho_{s} \mapsto \sqrt{\rm  {M}}_{i}\rho_{s}\sqrt{{\rm  {M}}_i}$, demonstrating that the state change   can be 
 upper-bounded with the probability of the outcome~\cite{Winter1999} as   $F(\rho_{s}, \tilde{\mathcal{J}}^{\mathcal{D}_{L}}_{i}(\rho_{s}))\geq p_{i} $ with    fidelity $F(\rho, \sigma)\equiv \large({\rm Tr}(\sqrt{\rho^{1/2}\sigma\rho^{1/2}})\large )^{2}$. 
This lemma finds applications in a wide range of topics, including the channel coding theorem~\cite{Winter1999}, state tomography~\cite{PRXQuantum, 10113, Chia20}, self-testing~\cite{PhysRevA.106.L010601}, differential privacy~\cite{10.1145/3313276.3316378}, the quantum OR problem~\cite{watts2022quantum}, quantum process learning~\cite{PRXQuantum.5.020367, PhysRevLett.129.160503}, quantum correlations~\cite{PhysRevX.13.041001, PhysRevLett.129.250504}, among others. Here, we demonstrate that the lemma can be established for intrinsic disturbance characterized by the SUR $\mathcal{J}^{\mathcal{N}}$ in terms of a family of such relations, which reduces to the form of  Winter's lemma when $\alpha = 1/2$.

The uncertainty and intrinsic disturbance effect are the characteristic properties of quantum measurements that  have been thought of as distinct and mainly studied separately in the common uncertainty relations and error-disturbance relations, respectively. Here, we show that they can be fundamentally related to each other by  showing that the uncertainty of a measurement, quantified by  $\alpha$-R\'{e}nyi entropy is  no less than its intrinsic disturbance effect, quantified by    $\alpha$-R\'{e}nyi divergence $\mathcal{D}_{\alpha}(\rho\|\sigma)$~\cite{PETZ198657}. Actually, they can lead to uncertainty relations. For instance, let $\alpha = 1$. We then have $\mathbf{H}(\mathbf{P}) \geq S(\rho\|\mathcal{J}^{\mathcal{N}}(\rho_{s}))$, which leads to the entropy uncertainty relation: $\mathbf{H}(\mathbf{P}_{A}) + \mathbf{H}(\mathbf{P}_{B}) \geq -\log c$ (see appendix),  where  $\mathbf{P}_{A}$ and $\mathbf{P}_{B}$ are the corresponding probability distributions obtained from independently measuring $A \equiv \{|i_{A}\rangle\}$ and $B \equiv \{|j_{B}\rangle\}$ on the state of interest and  $c \equiv \max_{i_{A}, j_{B}} |\langle i_{A}|j_{B}\rangle|^{2}$.    
Indeed, many other relations of this kind can be established similarly; they provide another method to derive preparation uncertainty relations and find also applications in quantum information science, for which, we leave a comprehensive investigation for future work. 

Information gained from a measurement can be quantified with different measures depending on the contexts. For instance, the operationally invariant measure specified as $\mathcal{I}(\mathbf{P})$  quantifies the information accumulated in each round to predict the outcome of a measurement.  By Eq.(\ref{invari}),  it can be seen that a non-trivial intrinsic disturbance implies information gain, demonstrating a complementary perspective to the idea that information gain implies disturbance. 

Groenewold introduced a measure of information gain as  the reduction of entropy for a provided instrument $\{\mathcal{J}_{i}\}$ as~\cite{Groenewold1971}:
\begin{eqnarray}
\mathcal{G}^{\mathcal{J}}(\rho_{s})\equiv S(\rho_{s})-\sum_{i}p_{i} S(\tilde{\mathcal{J}}_{i}(\rho_{s})).
\end{eqnarray}
This quantity depends on the employed instrument, and only $\mathcal{J}^{\mathcal{D}_{L}}$  always leads to a non-negative information gain~\cite{Ozawa1986OnIG}. By Eq.(\ref{compet}).  We consider this instrument and provide a competitive relationship between randomness and information gain, with their sum equating to the entropy of the measurement. Specifically, when the measured state is pure, and consequently $\mathcal{J}_{i}^{\mathcal{D}_{L}}(\rho_{s})$ is also pure, the information gain vanishes, and the measurement entropy becomes equal to the randomness generated.  Using Eq.(\ref{rand}), we also have $\mathcal{G}^{\mathcal{D}_{L}}+S(\rho_{s}\|\mathcal{J}^{\mathcal{N}}(\rho_{s})) \leq     {\rm H}(\mathbf{P}).$
This reveals a balance between $\mathcal{G}^{\mathcal{D}_{L}}$ and $S(\rho_{s}\|\mathcal{J}^{\mathcal{N}}(\rho_{s}))$ instead of a lower-bound on disturbance exploited in error-disturbance relation.

In the above considerations, we find $\mathcal{J}^{\mathcal{N}}$ is particularly straightforward to handle, as one can alternatively analyze a joint PVM and the reduction of a subsystem, enabling one to circumvent the common difficulty arising from the non-orthogonality of measurement elements, offering significant practical convenience.



\section{Conclusion} 
In this work, we tackle a long-standing open question: how to  characterize intrinsic disturbance in general measurement processes. By adopting a perspective---that a general measurement  can be viewed as a reduction of an extended projective measurement performed on the ancillary and the measured system, we derive a state updating rule for the concerned measurement as a reduction of the postulate of projection measurement. Our rule explicitly accounts for the disturbance effects attributed to  measurement elements and experimental realizations. Notably, it naturally recovers the projection postulate in the special case of projective measurements, thereby providing a unified theoretical foundation. We further apply this perspective to sequential measurement schemes, where we  take  the information carried by the ancillary system into account and  revisit key concepts in quantum measurement theory of  joint measurability and non-disturbance. Our  state-updating rule is also suitable for establishing quantitative connections between intrinsic disturbance and other fundamental quantum features, such as genuine randomness, quantum uncertainty, and information gain. In conclusion,  our framework thus offers a new foundation for understanding concepts related to quantum measurement and pave the way for exploring the constraints of disturbance effects in quantum information science.

\bibliography{pv} 

\newpage
\appendix
\onecolumngrid
\section{I. A brief review of quantum measurement theory}\label{section: A}
In this section, we begin with a brief overview of quantum theory, followed by the proof of Theorem 1. We will also explore the concepts of disturbance and the repeatability of measurement. Henceforth, we will employ the notions used in the main text: $\rho_{sa}\equiv \rho_{s}\otimes \rho_{a}$, $\rho_{sa|u}\equiv {\rm U}\rho_{sa}{\rm U}^{\dagger}$, and ${\bf Q}_{\rm H}\equiv \{{\rm Q}_{i|{\rm  H}}={\rm U}^{\dagger }{\rm Q}_{i}{\rm U}\}_{i}$. 

\subsection{A. Instruments of measurement}
 We make a brief review of measurement theory.   A general measurement can be described with positive operator-valued measures (POVM). If not projective, a POVM has to be extended to PVM by Naimark's dilation theorem.  
\begin{Theorem}[Naimark's dilation theorem]
For any POVM $\bf M$ performed on $\rho_{s}\in\mathcal{H}_{s}$, there are larger Hilbert's spaces $\mathcal{H}_{D}$ and  isometries $\rm V$ from  $\mathcal{H}_{s} $ to 
 $\mathcal{H}_{D}$ and  projective measurements $\mathbf{Q}\in \mathcal{H}_{D}$ such that 
                         ${\rm  {M}}_{i}=V^{\dagger} {\rm  {Q}}_{i}V, \forall i. $
\end{Theorem}
Without loss of generality, an isometry $V$  can be seen as the operations of adding an ancillary system in state $\rho_{a}$  initialized in $|0\rangle_{a}$  and acting a unitary operation $\rm U$, namely,  $ V: \rho_{s}  \mapsto {\rm U}\rho_{sa} {\rm U}^{\dagger}=\rho_{sa|u}$, and measurement is implemented  as  ${\rm Tr}(\rho_{s}{\rm M}_{i})={\rm Tr}(\rho_{s}V^{\dagger} {\rm  {Q}}_{i}V)={\rm Tr}(V\rho_{s}V^{\dagger}{\rm  {Q}}_{i})={\rm Tr}(\rho_{sa|u} {\rm  {Q}}_{i})$.   In textbooks~\cite{nielsen2001quantum, ref1}, an implementation  of measurement is often described as a dynamic process, where the measured state $\rho_{s}$ is coupled with an ancillary system represented by $|0\rangle \langle 0|_{a}$ through a unitary evolution ${\rm U}$, resulting in a joint state $\rho_{s} \otimes |0\rangle \langle 0|_{a}$. Subsequently, a PVM $\{|i\rangle \langle i|_{a}\}_{i}$ is performed on the ancillary system. In this context, the extended PVM $\mathbf{Q}$ takes the form $\{{\rm Q}_{i}\equiv \openone_{s} \otimes |i\rangle \langle i|_{a}\}$.  After tracing out the ancillary system, the measured state is given by ${\rm Tr}_{a}({\rm Q}_{i} {\rm U} \rho_{s} \otimes \rho_{a} {\rm U}^{\dagger} {\rm Q}_{i})$ (up to normalization), and SUR  is expressed as:
$$\mathcal{J}^{\mathcal{D}}_{i}: \rho_{s} \mapsto {\rm U}_{i|s} \sqrt{{\rm M}_{i}} \rho_{s} \sqrt{{\rm M}_{i}} {\rm U}^{\dagger}_{i|s},$$
where ${\rm U}_{i|s}$, determined by ${\rm U}$, is not essential for the measurement realization while altering the measured state additionally. Specifically, if ${\rm U}$ is chosen such that all $\{{\rm U}_{i|s}\}_{i}$ are identities, the SUR simplifies to:
$$\mathcal{J}^{{\mathcal{D}_{L}}}_{i}(\cdot): \rho_{s} \mapsto \sqrt{{\rm M}}_{i} \rho_{s} \sqrt{{\rm M}}_{i},$$
which is referred to as L\"{u}ders' rule and  regarded as a generalization of the postulate of PVMs to POVMs.

Such dynamic process of  a measurement can be seen from a Heisenberg picture  as  performing measurement ${\bf Q}_{\rm H}\equiv \{{\rm Q}_{i|\rm H}={\rm U}^{\dagger}\openone_{s} \otimes |i\rangle \langle i|_{a} {\rm U} \}$ on the joint state $\rho_{s}\otimes |0\rangle \langle 0|_{a}$. Then, we have the post-measurement state transformed according to $\mathcal{J}^{\mathcal{N}}$,    
\begin{equation}
\mathcal {J}^{\mathcal{N}}_{i}(\cdot): \rho_s \mapsto 
{\rm Tr}_{a}({\rm Q}_{i|\rm H}\rho_{sa}{\rm Q}_{i|\rm H} ).  
\end{equation}
Here, we  compare $\mathcal{J}^{\mathcal{D}}$ and $\mathcal{J}^{\mathcal{N}}$. First, we note that they are not trivially equivalent. The residual freedom $\{{\rm U}_{i|s}\}_{i}$ present in $\mathcal{J}^{\mathcal{D}}$ does not appear in $\mathcal{J}^{\mathcal{N}}$. To demonstrate this, we express ${\rm U}$ as ${\rm U}_{s}{\rm U}_{sa}$, where ${\rm U}_{s} \equiv \sum_{i} {\rm U}_{i|s} \otimes |i \rangle \langle i|_{a}$ and   ${\rm U}_{sa}|\psi\rangle |0\rangle_{a} = \sum_{i}\sqrt{{\rm M}_{i}}|\psi\rangle |i\rangle$ by definition.  Then we have 
\[
 {\rm Q}_{i|{\rm H}}={\rm U}^{\dagger} {\rm Q}_{i} {\rm U}= {\rm U}^{\dagger}_{sa} {\rm U}^{\dagger}_{s}  {\rm Q}_{i} {\rm U}_{s} {\rm U}_{sa} = {\rm U}^{\dagger}_{sa}  {\rm Q}_{i}{\rm U}_{sa},
\]
where we have used the  relation $[{\rm U}_{s}, {\rm Q}_{i}] = 0$.  

Second, we give the $\mathcal{J}^{\mathcal{N}}$ relevant to a provided  $\mathcal{J}^{\mathcal{D}}$ in terms of $\{{\rm U}, |0\rangle \langle 0|_{a}\}$. By expressing $\rm U$ as  ${\rm U}^\dagger=\sum_{ij}E^{\dagger}_{ij}\otimes |i\rangle\langle j|_{a} $, we have 
\begin{eqnarray}
&&V: \rho_{s}\mapsto  \rho_{s}\otimes | 0\rangle\langle 0|_{a},\nonumber \\
&& \textstyle {\rm Q}_{i|{\rm H}}={\rm U}^\dagger \openone_{s}\otimes|i\rangle\langle i|_{a}{\rm U}= (\sum_{j}E^{\dagger}_{ji})( \sum_{k}E_{ik})\otimes|j\rangle \langle k|_{a},\nonumber\\
&&{\rm M}_i=\tr_a ({\rm Q}_i{\rm I}_s\otimes |0\rangle\langle 0|_{a})=E^{\dagger}_{i0}E_{i0}, \nonumber\\
&&{\rm N}_{l|i}=\langle 0|{\rm Q}_{i}|l\rangle=E^{\dagger}_{0i}  E_{il}. 
\end{eqnarray}

We now apply our framework to the Ivanovic-Dieks-Peres (IDP) measurement and consider the optimal measurement for the unambiguous discrimination of two states $\{\cos\theta|0\rangle \pm \sin\theta|1\rangle\}$, which corresponds to a 3-outcome POVM on a qubit. For the sake of brevity we use notion  $\tan\theta=\cos\beta$, 
$${\rm M}_0=\sin^{2}\beta|0\rangle\langle0|,\quad {\rm M}_{1,2}={|\psi_\pm\rangle\langle\psi_\pm|} ,\quad |\psi_\pm \rangle=\frac{\tan\theta|0\rangle\pm|1\rangle}{\sqrt2}$$
 To realize this measurement,  one can choose the unitary $\rm U$ acting  on $\mathcal{H}_2\otimes \mathcal{H}_3$ as
$${\rm U}^\dagger=\begin{pmatrix}
\sin\beta&0&\frac{\cos\beta}{\sqrt2}&0&\frac{\cos\beta}{\sqrt2}&0&\\
0&0&\frac1{\sqrt2}&0&-\frac1{\sqrt2}&0\\
0&1&0&0&0&0\\
-\cos\beta&0&\frac{\sin\beta}{\sqrt2}&0&\frac{\sin\beta}{\sqrt2}&0\\
0&0&0&1&0&0\\
0&0&0&0&0&1\end{pmatrix}=\begin{pmatrix}E_{00}&E_{01}&E_{02}\\E_{10}&E_{11}&E_{12}\\
E_{20}&E_{21}&E_{22}\end{pmatrix}=\sum_{j,k=1}^2E^{\dagger}_{jk}\otimes|j\rangle\langle k|_{a},$$
and with projective measurement
${\rm Q}_{i|{\rm H}}={\rm U}^\dagger {\openone }\otimes|i\rangle\langle i|_{a}{\rm U}$.
The post-measurement states $\mathcal{J}_{i}^{\mathcal{N}}(\rho_{s})=\tr_a( {\rm Q}_{i|{\rm H}} \rho_{s}\otimes |0\rangle\langle 0|{\rm Q}_{i|{\rm H}})$ assume the forms
\begin{eqnarray}
&&\mathcal{J}^{\mathcal{N}}_{0}(\rho_{s})=\sin^2\beta\langle0|\rho_{s}|0\rangle\big(\sin^2\beta|0\rangle\langle 0|+\cos^2\beta|1\rangle\langle1|\big),\\
&&\mathcal{J}^{\mathcal{N}}_{1}(\rho_{s})=\langle\psi_+|\rho_{s}|\psi_+\rangle\left(|\psi_+\rangle\langle\psi_+|+\frac{\sin^2\beta}2|1\rangle\langle1|\right),\\
&&\mathcal{J}^{\mathcal{N}}_{2}(\rho_{s})=\langle\psi_-|\rho_{s}|\psi_-\rangle\left(|\psi_-\rangle\langle\psi_-|+\frac{\sin^2\beta}2|1\rangle\langle1|\right).
\end{eqnarray}
In  Ivanovic-Dieks-Peres (IDP) measurement,   the ancilla is a qubit instead of qutrit so that the diluted Hilbert space is a two-qubit system ${\mathcal H}_2\otimes {\mathcal H}_2$. The whole system undergoes a unitary evolution
$${\rm U}^\dagger=\begin{pmatrix}
\sin\beta&0&\frac{\cos\beta}{\sqrt2}&\frac{\cos\beta}{\sqrt2}&\\
0&0&\frac1{\sqrt2}&-\frac1{\sqrt2}\\
0&1&0&0\\
-\cos\beta&0&\frac{\sin\beta}{\sqrt2}&\frac{\sin\beta}{\sqrt2}\\
\end{pmatrix}\sim\begin{pmatrix}
\sin\beta&0&0&{\cos\beta}&\\
0&0&1&0\\
0&1&0&0\\
-\cos\beta&0&0&{\sin\beta},
\end{pmatrix}$$and
\begin{eqnarray}
{\rm Q}'_0={\rm U}^\dagger \openone_{s}\otimes|0\rangle\langle 0|_{a}{\rm U},\quad {\rm Q}'_j={\rm U}^\dagger |i\rangle\langle i|_{s}\otimes|1\rangle\langle 1|_{a}{\rm U} 
\end{eqnarray}
It is easy to see  ${\rm M}_i=\tr_a[{\rm Q}_i(\openone_{s}\otimes |0\rangle\langle 0|_{a})]$. The post-measurement states remain consistent with those obtained in the previous dilation. As demonstrated in the above example, our framework maintains its validity even when the dimension of the ancillary system does not correspond to the number of measurement outcomes. This flexibility is due to that one can extend the Hilbert space $\mathcal{H}_{2} \otimes \mathcal{H}_{2}$ to $\mathcal{H}_{2} \otimes \mathcal{H}_{3}$ through the addition of supplementary dimensions, followed by the application of a unitary transformation that converts the operators $\{{\rm Q}'_{i}\}_{i}$ into the desired form $\{{\rm Q}_{i|{\rm H}}\}_{i}$.

\subsection{B. The proof of  Eq.\ref{sur} in Theorem.1}

{\bf Theorem} [State updating rule]
For the realization of POVM by performing a joint PVM $\{{\rm Q}_{i}\}_{i}$  on state $\rho_{sa}$, the measured state is updated as  ${\rm Tr}_{a}({\rm Q}_{i}\rho_{sa} {\rm Q}_{i})$  with SUR  assuming  the form 
\begin{equation}
\mathcal{J}^{\mathcal{N}}_{i}:  \rho_s \mapsto 
{\rm  {M}}_{i}\rho_{s} {\rm  {M}}_{i}+\sum_{l\not=0} {\rm N}_{l|i}\rho_{s} {\rm N}^{\dagger}_{l|i},
\end{equation}
for  a set of operators  $\{  {\rm N}_{i|l}\}_{i,l}$ satisfying 
\begin{equation}
\sum_{l\not=0}{ {\rm N}}_{l|i}^{{\dagger}}{ {\rm N}}_{l|j}=\delta_{ij}{\rm  {M}}_{i}-{\rm  {M}}_{i}{\rm  {M}}_{j}.
\end{equation}
\begin{proof}
Without loss of generality, we can let the  ancilla be prepared in a pure state $|0\rangle_{a}$ where the  as any mixed state can be purified via a further extension.  In a basis $\{|i\rangle_a\}$ for the ancilla that includes the initial state as a member, a projective ideal measurement $\{{\rm  {Q}}_i\}$ in the composite system can be expanded
${\rm  {Q}}_{i}=\sum_{m,n}\Gamma^i_{mn}\otimes |m\rangle\langle n|_a$
with $\Gamma_{nm}^{i\dagger}=\Gamma_{mn}^i$ as ${\rm Q}_i^\dagger={\rm Q}_i$ and
$\sum_l \Gamma_{ml}^i\Gamma_{ln}^j=\delta_{ij}\Gamma_{mn}^i$
as ${\rm Q}_i{\rm Q}_j=\delta_{ij}{\rm Q}_i$.
To implement measurement $\mathcal{M}$, we  require ${\rm{Tr}}_{a}({\rm  {Q}}_{i}|0\rangle \langle 0|_a)={\rm  {M}}_{i}$ for all outcome $i$, from which it follows 
$\Gamma^i_{00}={\rm M}_i$. By denoting 
${\rm N}_{l|i}=\Gamma_{l0}^i$ for $l\not=0$, we have ${\rm N}_{l|i}=\Gamma^{i}_{l0}=\Gamma^{\dagger}_{0l}$ and ${\rm N}_{0|i}={\rm M}_{i}$.
We can see that each dilation ${\bf Q}$ gives one set $\{{\rm N}_{l|i}\}_{l, i}$ satisfying condition Eq(\ref{cond}). Conversely,  one set $\{{\rm N}_{i|l}\}$ satisfying condition Eq.(\ref{cond})  implies one such $ {\rm \bf Q}$ to realize the POVMs.  According to the postulate  for PVM on the total system, we obtain the final state of the system for outcome $i$ as
\begin{eqnarray}
 {\rm Tr}_{a}({\rm{Q}}_{i}\rho_{s}\otimes|0\rangle\langle 0|_{a}{\rm  {Q}}_{i})
=\textstyle {\rm  {M}}_{i}\rho_{s}{\rm{M}}_{i}+\sum_{l}{\rm \Gamma}^{i}_{l0}\rho_{s}{\rm \Gamma}^{i}_{0l},\\
= \textstyle {\rm  {M}}_{i}\rho_{s}{\rm{M}}_{i}+\sum_{l}{\rm N}_{l|i}\rho_{s}{\rm N}_{l|i}^{\dagger}
\end{eqnarray} 
for the system after obtaining outcome $i$.  
\end{proof}
\subsection{C. Proof of  Eq.\ref{cond} in Theorem.1} 
 For any channel with Kraus operators $\{{\rm N}_{l|i}\}$ with  ${\rm N}_{0i}={\rm M}_i$  satisfying Eq.(\ref{cond}), 
 namely, $\sum_{l\not=0}{\rm N}^\dagger_{l|j}{\rm N}_{l|i}=\delta_{ji}{\rm M}_i-{\rm M}_j {\rm M}_{i}$,  one can construct a PVM that reduces to the desired POVM. 

\begin{proof}
Let $\{|\psi_{k|i}\rangle\}_k$ be the eigenvectors of ${\rm M}_i$ corresponding to eigenvalues $\lambda_{k|i}\not=0$, namely,  ${\rm M}_{i}|\psi_{k|i}\rangle= \lambda_{k|i}|\psi_{k|i}\rangle$ and $\langle \psi_{k'|i} |\psi_{k|i}\rangle=\delta_{k k'} $. Define 
$$|q_{k|i}\rangle=\frac{1}{\sqrt{\lambda_{k|i}}}\sum_l {\rm N}_{l|i}|\psi_{k|i}\rangle\otimes|l\rangle_{a}$$
which are orthonormal
\begin{eqnarray}
 \langle q_{k'|j}|q_{k|i}\rangle&=&\frac{1}{\sqrt{\lambda_{k|i}\lambda_{k'|j}}}\langle\psi_{k'|j}|\sum_{l} {\rm N}^\dagger_{l|j}{\rm N}_{l|i}|\psi_{k|i}\rangle\\
&=&\frac{1}{\sqrt{\lambda_{k|i}\lambda_{k'|j}}}\langle\psi_{k'|j}| 
{\rm M}_j{\rm M}_{i}+\delta_{ji}{\rm M}_i-{\rm M}_j {\rm M}_{i}
|\psi_{k|i}\rangle
 =\delta_{ij}\delta_{kk'}
\end{eqnarray} 
so that we can define a projective measurement in the Hilbert space $\mathcal H_D=\mathcal{H}_s\otimes \mathcal{H}_a$
$${\rm Q}_i=\sum_{l,l'}{\rm N}_{l|i}\frac1{{\rm M}_i}{\rm N}_{l'|i}^\dagger\otimes |l\rangle\langle l'|_{a},$$
where we have denoted by
$$\frac1{{\rm M}_i}=\sum_{k,\lambda_{k|i}\not=0}\frac1{\lambda_{k|i}}|\psi_{k|i}\rangle\langle\psi_{k|i}|.$$
the generalized inverse of ${\rm M}_i$. One can check that  ${\rm Tr}({\rm Q}_{i}{\rm Q}_{j})=\delta_{ij} {\rm Q}_{i}$

Now we calculate
$$\tr_a({\rm Q}_i\openone_{s}\otimes|0\rangle\langle 0|_{a})={\rm N}_{0|i}\frac1{M_i}{\rm N}^\dagger_{0|i}=M_i$$
and
\begin{eqnarray*}
{\rm Q}_i(\rho\otimes|0\rangle\langle 0|){\rm Q}_{i}&=&\sum_{l,l'}{\rm N}_{l|i}\frac1{{\rm M}_i}
{\rm N}_{0|i}^\dagger\; \rho\;{\rm N}_{0|i}\frac1{M_i}{\rm N}_{l'|i}^\dagger\otimes|l\rangle\langle l'|_{a}\\
&=&\sum_{l,l'}{\rm N}_{l|i}\; \rho\; {\rm N}_{l'|i}^\dagger\otimes|l\rangle\langle l'|_{a}
\end{eqnarray*}
giving the desired channels $\tr_a[{\rm Q}_i(\rho_{s}\otimes|0\rangle\langle 0|_{a}){\rm Q}_{i}]=\mathcal {J}^{\mathcal{N}}_i(\rho_{s})$ after partial trace over ancilla. As ${\rm Q}=\sum_i {\rm Q}_i$ is a projection which might not span the whole space we can always complete it to a  basis  by adding to it the bases in the complement $1-{\rm Q}=\sum_\alpha|\phi_\alpha\rangle\langle\phi_\alpha|$.
\end{proof}

\subsection{D.  Disturbance in the measured state.}

The concept of intrinsic disturbance in POVMs has sparked significant debate in quantum measurement research~\cite{RevModPhys.86.1261, e21020142}. This contrasts  with PVMs, where intrinsic disturbance is explicitly  quantified by the distance $\mathcal{D}(\rho_{s}, \mathcal{J}^{\mathcal{P}}(\rho_{s}))$ between the initial state $\rho_{s}$ and the post-measurement state $\mathcal{J}^{\mathcal{P}}(\rho_{s})$.  A general measurement instrument can be described using a common formal structure:
\begin{equation}
\mathcal{J}_{i}(\cdot): \rho_s \mapsto {\rm Tr}_{a}\left({\rm U}_{i}{\rm Q}_{i}\rho_{sa}{\rm Q}_{i} {\rm U}_{i}^{\dagger}\right). 
\end{equation}
That is, performing  $\{{\rm Q}_{i}\}_{i}$ on the joint system $\rho_{sa}$, followed by a unitary operation $\rm U$ acting on the joint system. The irreversible state change is due to the projection measurement in this process $\rho_{sa}\mapsto {\rm Q}_{i}\rho_{sa}{\rm Q}_{i}$, which manifests in the measured state as
\begin{equation}
\mathcal{J}^{\mathcal{N}}_{i}(\cdot): \rho_s \mapsto {\rm Tr}_{a}\left({\rm Q}_{i}\rho_{sa}{\rm Q}_{i}\right).  \label{general2}
\end{equation}

Here, a few observations are worth noting. First, a given general channel $\mathcal{J}$ may correspond to different tuples $\{{\rm Q}_{i}, \rho_{a}\}$, leading to different $\mathcal{J}^{\mathcal{N}}$. This is reasonable because $\mathbf{Q}$ and $\rho_{s}$ are tied to the actual measurement process where different $\bf Q$ could lead to different kinds of the reduction of joint state,  and the reduced channel $\mathcal{J}$ for the measured system may not uniquely determine the real measurement process. Thus, even $\mathcal{J}$ is given, the channel characterizing intrinsic disturbance $\mathcal{J}^{\mathcal{N}}$ may vary, however, all the possible $\mathcal{J}^{\mathcal{N}}$ hold our arguments established in the paper.

To illustrate the reasonableness of our definition of the intrinsic disturbance in terms of distance between $\rho_{s}$ and $\mathcal{J}^{\mathcal{N}}(\rho_{s})$, let us consider the common instrument $\mathcal{J}^{\mathcal{D}}$, where a joint projective measurement (PVM) $\mathbf{Q} \equiv \{\openone_{s} \otimes |i\rangle \langle i|_{a}\}_{i}$ is performed on the state $\rho_{sa|u}$. The intrinsic disturbance on the joint measured state is quantified by $\mathcal{D}(\rho_{sa|u}, \mathcal{J}^{\mathcal{P}}(\rho_{sa|u}))$, and we then have
\begin{eqnarray}
    \mathcal{D}\left(\rho_{sa|u}, \sum_{i} {\rm Q}_{i} \rho_{sa|u} {\rm Q}_{i}\right) = \mathcal{D}\left(\rho_{sa}, \sum_{i} {\rm Q}_{i|{\rm H}} \rho_{sa|u} {\rm Q}_{i|{\rm H}}\right) \geq \mathcal{D}(\rho_{s}, \mathcal{J}^{\mathcal{N}}(\rho_{s})).
\end{eqnarray}
where  the unitary invariance of the state distance and data processing inequality have been used.

Here, we would like to highlight that for a given POVM, it is possible to estimate the intrinsic disturbance $\mathcal{D}(\rho_{s}, \mathcal{J}^{\mathcal{N}}(\rho_{s}))$ to some extent, even without knowledge of the dilation. We define $\mathcal{J}^{\mathcal{N}}(\rho_{s})=\lambda \rho^{\eta}_{s}+\bar{\lambda} \rho^{\bar{\eta}}_{s}$, where $\lambda \rho^{\eta}_{s} \equiv \sum_{i} {\rm M}_{i} \rho_{s} {\rm M}_{i}$, $\bar{\lambda} \rho^{\bar{\eta}}_{s} \equiv \mathcal{J}(\rho_{s}) - \rho^{\eta}_{s}$, and $\lambda \equiv {\rm Tr} \left(\sum_{i} {\rm M}_{i} \rho_{s} {\rm M}_{i}\right)$. We find that  $\lambda     \mathcal{D}(\rho_{s}, \rho^{\eta}_{s}) + \bar{\lambda}     \mathcal{D}(\rho_{s}, \rho^{\bar{\eta}}_{s}) \geq     \mathcal{D}(\rho_{s}, \lambda \rho^{\eta}_{s} + \bar{\lambda} \rho^{\bar{\eta}}_{s}) \geq \lambda     \mathcal{D}(\rho_{s}, \rho^{\eta}_{s}) - \bar{\lambda}     \mathcal{D}(\rho_{s}, \rho^{\bar{\eta}}_{s}),$ where we assume that the state distance satisfies the triangle inequality, which is typically a property of geometric distance measures  such as trace distance. Consequently, we can establish lower and upper bounds:
$\lambda \mathcal{D}(\rho_{s}, \rho^{\eta}_{s}) \pm \bar{\lambda} \mathcal{D}_{0},$ where  $\mathcal{D}_{0} \equiv \max_{\varrho} \mathcal{D}(\rho_{s}, \varrho)$.

\subsection{E. Conceptual comparison of PVM and  POVM}

 While a POVM can be viewed as a reduction of some PVM, it does not share the same physical meaning as the joint PVM. This is because a POVM can arise from dilated PVMs with different physical interpretations, some of which may be specific to a joint system and not well-defined for a subsystem, such as parity measurements.   In the realization of a POVM, repeating the joint PVM $\mathbf{Q}$ does not imply repeating the relevant POVMs on the subsystem due to the update of the ancillary system state. Consequently, POVMs do not inherently assume repeatability, where  a measurement is said to be repeatable if repeating it consistently yields the same outcome from the last round~\cite{Carmeli2007, PhysRevLett.92.070403}.

To demonstrate this point, we examine a seven-outcome POVM defined as $\{a_{i} |\phi_{i}\rangle_{s} \langle \phi_{i}|\}_{i}$, where the coefficients $a_{i}$ are given by $a_{i} = \frac{2}{3}$ for $i = 2, 3$ and $a_{i} = \frac{1}{3}$ for all other values of $i$.
 \begin{eqnarray}
\centering
\begin{tabular}{l  l  l}  
$|\phi_{1}\rangle_{s}=|1\rangle_{s}$,&  $|\phi_{2}\rangle_{s}=|0\rangle_{s}$,& $|\phi_{3}\rangle_{s}=|2\rangle_{s}$,\\
 $|\phi_{4}\rangle_{s}=|1+2\rangle_{s}$,& $|\phi_{5}\rangle_{s}=|1-2\rangle_{s}$,&$|\phi_{6}\rangle=|0+1\rangle_{s}$,\\	$|\phi_{7}\rangle_{s}=|0-1\rangle_{s}$,& \quad & \quad \\     
\end{tabular}
\end{eqnarray}
where we have used the notions $|i\pm j\rangle=\frac{1}{\sqrt{2}}(|i\rangle\pm |j\rangle)$.  
Here one can  construct $\mathbf{Q}$   using qutrit  ancillary system  
 \begin{eqnarray}\nonumber
\centering
\begin{tabular}{l  l  l}  
$|\Phi_{1}\rangle=|1\rangle_{s}|1\rangle_{a}$,&  $|\Phi_{2}\rangle=|0\rangle_{s}|0+1\rangle_{a}$,& $|\Phi_{3}\rangle=|0\rangle_{s}|0-1\rangle_{a}$,\\ 
 $|\Phi_{4}\rangle=|2\rangle_{s}|1+2\rangle_{a}$,& $|\Phi_{5}\rangle=|2\rangle_{s}|1-2\rangle_{a}$,&	$|\Phi_{6}\rangle=|1+2\rangle_{s}|0\rangle_{a}$,\\	$|\Phi_{7}\rangle=|1-2\rangle_{s}|0\rangle_{a}$,& $|\Phi_{8}\rangle=|0+1\rangle_{s}|2\rangle_{a}$,& $|\Phi_{9}\rangle=|0-1\rangle_{s}|2\rangle_{a}$\\     
\end{tabular}
\end{eqnarray}
Let ${\rm Q}_{i}=|\Phi_{i}\rangle \langle \Phi_{i}|$ (the measurement used to demonstrate nonlocality without entanglement~\cite{bennett1999quantum}),  and  $\rho_{a}= {\openone}_{a}/3$,  we have  $ {\rm  {M}}_{1}={\rm Tr}_{a}(\rho_{a} {\rm Q}_{1})$,  $ {\rm  {M}}_{2}={\rm Tr}_{a}(\rho_{a} ({\rm Q}_{2}+{\rm Q}_{3}))$,   $ {\rm  {M}}_{3}={\rm Tr}_{a}(\rho_{a} ({\rm Q}_{4}+{\rm Q}_{5}))$,   and $ {\rm  {M}}_{i+2}={\rm Tr}_{a}(\rho_{a} {\rm Q}_{i+2})$, for $i\geq 2$. Let measurement outcome, for instance,  be one, the post-measurement state of the joint system is $|0\rangle_{s}|0\rangle_{a} $, repeating the PVM  $\mathbf{Q}$  the relevant  to  POVM measured in the second round, given by  $ {\rm  {M}}'_{i}\equiv {\rm Tr}(|0\rangle\langle 0|_{a} {\rm Q}_{i})$, 
are ${\rm  {M}}'_{1}=|1\rangle_{s} \langle 1|, {\rm  {M}}'_{2}={\rm  {M}}'_{3}=1/2|0\rangle \langle 0|_{s},  {\rm  {M}}'_{4}={\rm  {M}}'_{5}=1/2|2\rangle\langle 2|_{s}$ and ${\rm  {M}}'_{i}=0$ for $i\geq 6$. This POVM  differs from POVM in the first round. Then,  repeated  PVMs realize different  POVMs.

\section{II. Relation between intrinsic disturbance and  notions relate to quantum measurement}

\subsection{A. Proof of Uncertainty-disturbance relation and gentle measurement lemma}
One enduring interest in the foundations of quantum mechanics is to understand the fundamental properties of quantum measurement, specifically uncertainty and disturbance, driven by both theoretical curiosity and practical applications. Since uncertainty and disturbance are inherent features of quantum measurements, one might wonder whether they can be related to one another. In the following discussion, we provide a positive answer to this question.

Here, the chosen state distance for defining disturbance is the R\'{e}nyi divergence.
$$
\mathcal{D}_\alpha(\rho\|\sigma)=
\begin{cases}
(\alpha-1)^{-1} \log\tr\left(\sigma^{(1-\alpha)/(2\alpha)} \rho \sigma^{(1-\alpha)/(2\alpha)} \right)^\alpha & \text{if}\ \alpha\in(0,1) {\rm or} (1,\infty) \quad {\rm  supp \rho}\subset {\rm supp \sigma}  \,, \\
\infty & \text{if} \ {\rm otherwise}
\end{cases}
$$
We  also have  $\mathcal{D}_{\alpha \to 1}=S(\rho\|\sigma)$.
We first assume  $\rho_{s}$ is a pure state and let $\rho_{a}=|0\rangle \langle 0|_{a}$ and so does $\rho_{sa}$, then $ {\rm Tr}([\mathcal{J}^{\mathcal{P}}_{i}(\rho_{sa})]^{\tau} \rho)=  {\rm Tr}[ ({\rm Q}_{i} \rho_{sa} {\rm Q}_{i} )^{\tau}\rho_{sa}] = p^{\tau +1}_{i}$. Note that, for a positive Hermitian operator $O$, $\operatorname{Tr}O^{\tau}\geq (\operatorname{Tr}O)^{^\tau} $ for $ 0< \tau < 1$ and $\operatorname{Tr}O^{\tau}\leq (\operatorname{Tr}O)^{^\tau}$ for $\tau\geq1$. Specify $\sum_{i}{\rm Q}_{i}\rho_{sa}{\rm Q}_{i}$ by $\tilde{\rho}_{sa}$ for brevity, we have
	\begin{eqnarray}\nonumber
	\operatorname{Tr}
	({\tilde\rho}^{\frac{1-\alpha}{2\alpha}}_{sa}\rho_{sa}
	{\tilde\rho}^{\frac{1-\alpha}{2\alpha}}_{sa})^{\alpha}&\geq&
	[\operatorname{Tr}({\tilde\rho}^{\frac{1-\alpha}{2\alpha}}_{sa}
	\rho_{sa}{\tilde\rho}^{\frac{1-\alpha}{2\alpha}}_{sa})]^{\alpha}=  [\operatorname{Tr}({\tilde\rho}^{\frac{1-\alpha}{\alpha}}_{sa}
	\rho_{sa})]^{\alpha}
	\nonumber\\=(\sum_{i}
	p^{\frac{1}{\alpha}}_{i})^{\alpha}&=&
	\|\mathbf{P}\|_{\frac{1}{\alpha}}, \quad \quad 0<\alpha<1,\nonumber\\
	\operatorname{Tr}
	({\tilde\rho}^{\frac{1-\alpha}{2\alpha}}_{sa}\rho_{sa}
	{\tilde\rho}_{sa}^{\frac{1-\alpha}{2\alpha}})^{\alpha}&\leq&
	\|\mathbf{P}\|_{\frac{1}{\alpha}}, \quad \quad \alpha>1. \nonumber
	\end{eqnarray}
	Using the notion of $\alpha$-R\'{e}nyi entropy ${\rm H}_{\alpha}(\bf{P})=\frac{\alpha}{1-\alpha}\log \|\mathbf{P}\|_{\alpha}$, we have  a unified expression as 
	\begin{eqnarray}
	{\rm H}_{\frac{1}{\alpha}}(\mathbf{P})\geq {\mathcal{D}}_{\alpha}(\rho_{sa}|\tilde \rho_{sa})\geq  {\mathcal{D}}_{\alpha}(\rho_{s}|\mathcal{J}^{\mathcal{N}} (\rho_{s})), \label{UDRre}
	\end{eqnarray}
Specifically, one may only  concern  the post-measurement state corresponding to a particular outcome \(i\). Following a similar approach to the derivation above, we focus on the region \(\alpha \in [1/2, 1)\) and assume \(\rho_{s}\) to be a pure state. Consequently, \(\rho_{sa} = \rho_{s} \otimes \rho_{a}\) is also a pure state, and the post-measurement state \(\tilde{\rho}_{i|sa}\) is given by \(\tilde{\rho}_{i|sa} = \frac{{\rm Q}_{i} \rho_{s} \otimes \rho_{a} {\rm Q}_{i}}{p_{i}}\).  We then have  $\operatorname{Tr}
	({\tilde\rho}^{\frac{1-\alpha}{2\alpha}}_{i|sa}\rho_{sa}
	{\tilde\rho}^{\frac{1-\alpha}{2\alpha}}_{i|sa})^{\alpha}=[\operatorname{Tr}({\tilde\rho}_{i|sa}^{\frac{1-\alpha}{2\alpha}}
	\rho{\tilde\rho}_{i|sa}^{\frac{1-\alpha}{2\alpha}})]^{\alpha}=[\operatorname{Tr}({\tilde\rho}_{i|sa}^{\frac{1-\alpha}{\alpha}}
	\rho_{sa})]^{\alpha}=[1/p_{i}{\rm Tr}({\rm Q}_{i} \rho_{sa} {\rm Q}_{i} \rho_{sa})]^{\alpha}=p^{\alpha}_{i}$ or written as 
	\begin{eqnarray}
	\alpha/(\alpha-1)\log p_{i} \geq    \mathcal{D}_{\alpha}(\rho_{sa} \|\tilde{\rho}_{i|sa} ) \geq \mathcal{D}_{\alpha}(\rho_{sa}\|\tilde{\mathcal{J}}^{\mathcal{N}}_{i} (\rho_{sa})), \label{udrgml}
	\end{eqnarray}  
When $\alpha=1/2$, we  have   $p^{1/2}_{i}=\operatorname{Tr}
	(\tilde{\rho}_{i|sa}^{\frac{1}{2}}\tilde{\rho}_{sa}
	{\tilde\rho}^{\frac{1}{2}}_{i|sa})^{1/2}=\sqrt{F(\rho_{sa}, \tilde \rho_{i|sa})}\leq \sqrt{{\rm F}(\rho_{s}, \mathcal{\tilde J}^{\mathcal{N}}_{i}(\rho_{s})})$ and  recover the GML by Winter as
\begin{eqnarray}
p_{i} \leq {\rm F}(\rho_{s}, \mathcal{\tilde J}^{\mathcal{N}}_{i}(\rho_{s})). 
	\end{eqnarray}
    In Eq. (\ref{udrgml}), the left-hand side is convex while the right-hand side is concave, indicating that this relation is applicable to mixed states as well.

We can derive the Maassen-Uffink uncertainty relation~\cite{PhysRevLett.60.1103}, given by 
\[
\mathbf{H}(\mathbf{P}_{A}) + \mathbf{H}(\mathbf{P}_{B}) \geq -\log c,
\]
where \(c \equiv \max_{ij} c_{ij}\), from the above uncertainty-disturbance relation. Setting \(\alpha = 1\), we find that 
\[
\mathbf{H}(\mathbf{P}_{A}) \geq S(\rho_{s} \| \mathcal{J}^{\mathcal{N}}(\rho_{s})).
\]

Let us denote the measurement operators as \(A \equiv \{|i_{A}\rangle\}\) and \(B \equiv \{|j_{B}\rangle\}\). Consequently, we can express 
$\mathcal{J}^{\mathcal{N}}(\rho_{s}) = \sum_{i} p_{i} |i_{A}\rangle \langle i_{A}|.$
When we perform measurement \(B\) on both \(\rho_{s}\) and \(\mathcal{J}^{\mathcal{N}}(\rho_{s})\), we obtain the probabilities 
${\bf P}_{B} \equiv\{ p_{j|B}=\mathrm{Tr}(\rho_{s} |j_{B}\rangle \langle j_{B}|)\}$
and 
${\bf P}'_{B} = \{P'_{j|B} = \mathrm{Tr}(\mathcal{J}^{\mathcal{N}}(\rho_{s}) |j_{B}\rangle \langle j_{B}|) = \sum_{i} c_{ij} p_{i|A}\}$ with $c_{ij}=|\langle j_{B}|i_{A} \rangle |^{2} $. 
By applying the data processing inequality, we have 
$\mathbf{H}(\mathbf{P}_{A}) \geq \mathrm{H}(\mathbf{P}_{B} \| \mathbf{P}'_{B})$.
This leads us to the following inequality:
$\mathrm{H}(\mathbf{P}_{B} \| \mathbf{P}'_{B}) \geq \sum_{j} p_{j|B} \log_{2} \frac{p_{j|B}}{p'_{j|B}} \geq -\mathrm{H}(\mathbf{P}_{B}) - \sum_{j} p_{j|B} \log_{2} \sum_{i} c_{ij} p_{i} \geq -\mathrm{H}(\mathbf{P}_{B}) - \log_{2} c,$
where \(p'_{j|B} = \sum_{i} c_{ij} p_{i} \leq c\).
Combining this with $\mathrm{H}(\mathbf{P}_{A}) \geq \mathrm{H}(\mathbf{P}_{B} \| \mathbf{P}'_{B}),$
we arrive at the desired uncertainty relation.
\[
\mathbf{H}(\mathbf{P}_{A}) + \mathbf{H}(\mathbf{P}_{B}) \geq -\log c.
\]
    
\subsection{B. Information-disturbance balance}
In considering the question: \emph{How much information is gained by performing a given quantum measurement?},  Groenewold introduced a measure of information gain as  the reduction of entropy for a provided instrument $\{\mathcal{J}_{i}\}_{i}$ as 
\begin{eqnarray} 
\mathcal{G}^{\mathcal{J}}\equiv S(\rho_{s})-\sum_{i}p_{i}S(\tilde{\mathcal{J}}_{i}(\rho_{s})). 
	\end{eqnarray}
This quantity is always positive for  $\mathcal{J}^{\mathcal{D}}$, for which,  we have  the relation between information gain $\mathcal{G}^{\mathcal{D}}$ and state disturbance quantified by $S(\rho\|\mathcal{J}^{\mathcal{N}}(\rho_{s}) )$ as
    \begin{eqnarray}
 \mathcal{G}^{\mathcal{D}}+  S(\rho_{s}\|\mathcal{J}^{\mathcal{N}}(\rho_{s}))\leq     {\rm H}(\mathbf{P}).
  \end{eqnarray}
\begin{proof}
    \begin{equation}
	\begin{aligned} S(\rho_{sa|u}\|\mathcal{J}^{\mathcal{P}}(\rho_{sa|u}))&=\operatorname{Tr}(\rho_{sa|u}(\log\rho_{sa|u}-\log {\mathcal{J}}^{\mathcal {P}}(\rho_{sa|u})))\nonumber \\
     &=-S(\rho_{sa|u})+S(\mathcal{J}^{\mathcal{P}}(\rho_{sa|u})) \nonumber \\
 &=-S(\rho_{sa})+\operatorname{H}(\mathbf{p}) +\sum_{i}p_{i}S(\tilde{\mathcal{J}}^{\mathcal {P}}(\rho_{sa|u})) \nonumber \\
  &=-S(\rho_{s})+\operatorname{H}(\mathbf{p}) +\sum_{i}p_{i}S(\tilde{\mathcal{J}}^{\mathcal{D}}(\rho_{s}))\nonumber \\
&=  -\mathcal{G}^{\mathcal{D}}+{\rm H}(\mathbf{P})\nonumber
 \nonumber
	\end{aligned}
	\end{equation}
    where  $\rho_{sa}=\rho_{s}\otimes |0\rangle \langle 0|_{a}$, $\rho_{sa|u}={\rm U} \rho_{sa} {\rm U}^{\dagger}$, $\mathcal{J}^{\mathcal{P}}(\rho_{sa})=\sum_{i}{\rm Q}_{i}\rho_{sa|u}{\rm Q}_{i}=\sum_{i}{\rm U}_{i|s} \sqrt{{\rm M}_{i}}\rho_{s} \sqrt{{\rm M}_{i}} {\rm U}^{\dagger}_{i|s} \otimes |i\rangle \langle i|_{a}=$$  \sum_{i}\mathcal{J}_{i}^{\mathcal{D}} (\rho_{s}) \otimes|i\rangle \langle i|_{a}=\sum_{i}p_{i}\tilde{\mathcal{J}}_{i}^{\mathcal{D}} (\rho_{s}) \otimes|i\rangle \langle i|_{a}$ thus  $S(\mathcal{J}^{\mathcal{P}}(\rho_{sa|u}))= {\mathbf{H}}(\mathbf{P})+\sum_{i}p_{i}S(\tilde{\mathcal{J}}_{i}^{\mathcal{D}}(\rho_{s}))$, and we have used  $S(\rho_{sa|u})=S(\rho_{sa})=S(\rho_{s})$. 
We note that $\mathcal{R}=S(\rho_{sa|u}\|\mathcal{J}^{\mathcal{P}}(\rho_{sa|u})) $, we also have 
    \begin{eqnarray}
 \mathcal{G}^{\mathcal{D}}+\mathcal{R} =  {\rm H}(\mathbf{P}).
  \end{eqnarray}    
Noting that $ S(\rho_{sa|u}\|\mathcal{J}^{\mathcal{P}}(\rho_{sa|u}))\geq S(\rho_{s}\| \mathcal{J}^{\mathcal{N}}(\rho_{s}))$, we then have the desired result
    \begin{eqnarray}
 \mathcal{G}^{\mathcal{D}}+S(\rho_{s}\|\mathcal{J}^{\mathcal{N}}(\rho_{s})) \leq  {\rm H}(\mathbf{P}).
  \end{eqnarray}

\end{proof}


\subsection{C. Disturbance lower-bounds information gain}
Here,  the operational  invariant information defined in~\cite{PhysRevLett.83.3354}  is $\mathcal{I} \equiv 1 -\|\mathbf{p}\|^{2}_{2}$, for which, we have   
\begin{eqnarray}
\mathcal{I}\geq \mathcal{D}^{2}_{\rm Tr}(\rho, \mathcal{J}^{\mathcal{N}}(\rho_{s})).
  \end{eqnarray}
For trace-distance $    \mathcal{D}_{\rm Tr}(\rho_{1}, \rho_{2})\equiv \frac12\rm{Tr}|\rho_{1}-\rho_{2}|$.

\begin{proof}
     Let the auxiliary system be in a pure state $\rho_{a}$   and so does $\rho_{s}$, we have
\begin{eqnarray}
\textstyle \sqrt{1-\|\mathbf{p}\|_{2}^{2}}&= &\textstyle  \sqrt{1-\sum_{i}p^2_{i}}
 = \sqrt{1-\operatorname{Tr}(\rho_{sa|u}\cdot \mathcal{J}^{\mathcal{P}}(\rho_{sa|u}))}\nonumber \\
&  \geq& \sqrt{1-\operatorname{F}(\rho_{sa|u},  \mathcal{J}^{\mathcal{P}}(\rho_{sa|u}))}= \sqrt{1-\operatorname{F}(\rho_{sa},  \sum_{i}{\rm Q}_{i|{\rm H}}\rho_{sa}{\rm Q}_{i|{\rm H}})} \\
&\ge &\textstyle \frac{1}{2}\operatorname{Tr}|\rho_{sa}- \sum_{i}{\rm Q}_{i|{\rm H}}\rho_{sa}{\rm Q}_{i|{\rm H}}|\geq \textstyle \frac{1}{2}\operatorname{Tr}|\rho_{s} - \mathcal{J}^{\mathcal{N}}({\rho}_{s})|\\
&=&\mathcal{D}_{\rm tr}(\rho_{s}, \mathcal{J}^{\mathcal{N}}(\rho_{s}))\nonumber
\end{eqnarray}
where we have used   $\operatorname{Tr}(\rho_{sa|u}  \sum_{i}{\rm Q}_{i} \rho_{sa|u} {\rm Q}_{i} )=\sum_ip_i^2$    in the second equality and  the  $\operatorname{Tr}(\rho_{1} \rho_{2})\leq \rm{F}(\rho_1,\rho_2) $ for any two state $\rho_{1}$ and $\rho_{2}$   in the first inequality and unitary invariance of state distance and the Fuchs-van de Graaf inequality  $\sqrt{1-\rm{F}(\rho_1,\rho_2)}\geq \textstyle \frac{1}{2}{\rm Tr}|\rho_{1}-\rho_{2}|$  in the second inequality and the data processing inequality in the last inequality. Our relation applies to mixed state as the left hand side is concave while the right-hand side is convex.

\end{proof}

\clearpage

\end{document}